# A STUDY OF METHODS FROM STATISTICAL MECHANICS APPLIED TO INCOME DISTRIBUTION

E. OLTEAN, F. V. KUSMARTSEV[*]

***Abstract.*** *Research developed so far in Econophysics was dedicated to the analysis and prediction of financial data, financial markets, and financial products using methods and laws from Physics. More recently, a new direction was developed towards the study of macroeconomic aggregates and in particular to the income distribution within the boundaries of a state. The purpose of our endeavour is to form a database regarding income from different countries both developed and underdeveloped, and apply methods from Statistical Mechanics to make the analysis of these data and to establish definite causes of macroeconomic evolutions and financial crisis, and establish analogy with physical phenomena. We tried to analyse the data from six countries: France, Finland, Italy, Romania, Italy, and Hong Kong by using very reliable sources such as National Institute of Statistics or National Bank in each case. The data used were the distribution of income per capita divided by deciles in two variants: mean income and upper limit on income measured on annual bases. Each national economy was assimilated to a grand canonical ensemble, while particles were considered individuals or households. Having made these assumptions, we tried Fermi-Dirac distribution, Bose-Einstein distribution, and occasionally Boltzmann-Gibbs distribution in order to determine which is optimal for income distribution. The best fit to the data was observed in the case of Fermi-Dirac distribution, for which the coefficient of determination showed the best goodness of fit to the data. Using this distribution for data (spun throughout more years), we obtained the underlying critical parameters of annual income distribution such as chemical potential and temperature. The next step was to explore the evolution of income using economic analogues to chemical potential and temperature. Using as background the Yakovenko's analogy between temperature from thermodynamic systems and nominal income from Economics, we found other analogies that would allow further analysis and explanation of income.*

***Keywords:*** *statistical mechanics, Bose-Einstein distribution, income distribution, Fermi-Dirac distribution*

[*] Department of Physics, Loughborough University, Leicestershire, LE11 3TU, UK, e-mail: elvis.oltean@alumni.lboro.ac.uk and f. kusmartsev@lboro.ac.uk

# Chapter I. Applications of Fermi-Dirac distribution to income distribution

## I.1. Introduction

The word *econophysics* was first used at an international workshop held in Calcutta in 1995, while the first book containing the word econophysics in its title is authored by Mantegna and Stanley [1] [2].

However, the systematic relationship between Economics and Physics commenced at least 130 years ago, when the Marginalist school began to make massive use of mathematics, borrowing their tools from Physics. Thus, Alfred Marshall and William Jevons, the main theoreticians of the Marginalist school, envisaged making Economics as a second-Physics, while the fundamental notion of utility was to be treated as a Mechanics of human interest. Implementation of this interdisciplinary field was necessary to overpass the limits of Economics, given that mechanical epistemology influenced to a large extent the modern economic thinking. The limitation was determined, primarily, by the fact that economic activity is influenced decisively by human behaviour and human interrelations, which change throughout the time, being impossible to explain it with mechanical-type methods and laws [3].

Econophysics emerged as a consequence of the application of methods from Statistical Physics to financial markets, but subsequent scientific works in Econophysics focussed on three areas. The first area is about prices on capital market, currency exchange rates, and the prices of goods. The second one is about size of firms, macroeconomic aggregates, individual income and wealth. The third one is about analysis of economic phenomena using network type models [4] [5].

## I.2. Short Literature Review and Theoretical Background

There are several trends which worth to be mentioned for a better understanding of the problems that come across the income distribution models. Most important, there is no model for an income distribution which would allow figuring out the features for the whole income range [6].

The income distribution seems to exhibit two regimes of behaviour. For low and middle income population, which is represented from 90% up to 99 % of the population, the income may show according to different authors and papers, lognormal, gamma, Boltzmann-Gibbs distribution or other exponential-like functions [7]. For higher income part of population, which accounts up to 10% of the entire population, the income exhibits a power law of Pareto type. The only distribution which is claimed to fit the entire range is Tsallis [8].

Authors are often confused about the meaning of between wealth and income. Wealth evolution can influence the income by way of assets prices which evolve throughout the time and the financial yield that they give. Income can add to the already existing wealth. Also, wealth is considered to be a stock while income is considered a flow [9].

Data are considered to be by many authors a big problem which reflects on the accuracy of their results. However, more and more reliable statistical data are provided.

Most of the authors did not tackle the evolution after the beginning of the crisis. The only exception is [8], which claims there are some changes in the income distribution for years when the recession started to affect to a large extent the Argentinian economy.

### I.3. Data Description

The data used for analysis is mean income and upper limit on ranges of income divided into deciles of population/households (equal bins of 10 % of the population/households). Thus, for France, Finland, and Italy we were able to get the data both for mean income and upper limit on income. The latter term for income is in accordance with the term used by National Institute of Statistics of Finland to describe the upper value of each income decile. For Romania and Mexico, we were able to obtain mean income, while for Hong Kong we used for the analysis mean monthly median income from different years. Also, the data were made available in different formats as follows: for France and Finland, the data were with regard to individuals, whereas for Romania, Mexico, Italy, and Hong Kong the data were about households. The data were considered in different monetary units. For example, in case of France they were expressed in euro for the entire time period considered. Italy was considered both for lire, which was national currency before euro (last year considered for lire in of Italy was in year 1998) and euro starting from year 2000. In the case of Finland, the data were altered such that the numerical value for each year income according to last year considered, making the data more reliable and realistic. In Romania, the data were expressed in leu which was the currency until July 2005, when a new currency was introduced called heavy leu. The ratio between 1 heavy leu = 10000 leu. In Mexico, the data were expressed in US dollars, making the data more accurate given the relatively low inflation in the USA. Finally, in Hong Kong the figures were expressed in Hong Kong dollars depending on the purchase power parity in the final year considered (2001), which is just another method (just like in case of Finland) to alter the data in order to make them more accurate and reliable for analysis purpose.

The income considered can be gross income which is the revenue obtained from different kind of sources such as wages, dividends, rents, and so forth. Generally, the income of 90 % of the population depend almost entirely on wages, whereas the income of upper 10 % (or sometimes even less) depend on the prices of the assets.

The income analysed can be also disposable income or net income which is defined as the income which an individual or household have available to spend or to save after taxes are paid and/or social benefits are received [10]. The National Institute of Statistics from France calls it income before redistribution.

*Net (Disposable) Income = Gross Income – Taxes + Social Benefits* (1)

Thus, most of the income data analysed in this paper is about net income, except a set of data regarding inactive people's income, income before redistribution, and mean wealth provided for France, which will be analysed as well.

The upper limit on income comprises 90% of the population, as for the upper 10 % data were not provided. In the case of mean income, the data cover the entire population as it is possible to calculate the mean income for the richest part of the population. For three countries (France, Finland, and Italy), we were able to get both upper limit on income and mean income which can lead to supplementary results. In case of France, more data were made available such as income distribution for non-active persons, mean wealth, and income before redistribution.

The data provided are very reliable especially for France, Italy, Finland, and Romania being provided by National of Statistics for most of them and in case of Italy by the National Bank. Besides, different bodies of the European Union double-check the accuracy and veracity of the data provided. In case of Mexico and Hong Kong the data where provided by different national bodies which analyse this.

### I.4. Methodology

The paper analyses the distribution of net income, gross income, and median income of the population according to allocations specific for a market economy in equilibrium, under similar conditions for thermal equilibrium in thermodynamics. The analogy with thermodynamics used here is grand-canonical ensemble, where both number of indivi-duals/particles and the amount of energy/money are hold constant but the average for both variables is the same for a longer period of time. The other two cases are micro-canonical ensemble (fixed number of parti-cles/individuals) and canonical ensemble (fixed number of particles/indi-viduals and variable amount of energy but the average of energy is fixed) which do not comply with reality of a national economy [11]. Grand-canonical ensemble is the best approximation for a national economy as population varies slowly over long period of time, so it can be approximated as constant. Also, the average is considered to be constant for a certain amount of time.

Fermi-Dirac distribution is the best model that can be applied considering that is highly unlikely for a person or a household to have exactly the same income, especially in the case of net income. While for gross income this is more likely (two people working in a public institution
– where wages are less flexible – working in a similar position with a similar background), for net income there other things that differentiate it among various people (such as taxation level, payment for different credits granted, different transfers from public budget).

Since all the countries analysed have populations of millions of persons and each person has virtually a different income within an interval of tens of thousands of monetary units of income, this can be approximated as a continuous interval. Subsequently:

$$n(x) = \frac{c}{\exp\frac{x-\mu}{T} + 1} \qquad (2)$$

where $n(x)$ is the number of fermionic particles within some energy interval or in economic terms is the number of individuals having an amount of money within some level of income.

We tried to apply Fermi-Dirac distribution and Bose-Einstein distribution to the data described before. The main criterion to judge on how applicable is each distribution was the coefficient of determination ($R^2$). Subsequently, we were able to determine that Fermi-Dirac distribution has a better fit to the data. Subsequently, in the following paper, we will present only the results from the application of Fermi-Dirac distribution.

## I.5. Data analysis

The preferred method for data description is cumulative method instead of normal method. This implies that for zero income the percentage of the population is 100%. Thus, the more the level of income increases the cumulative percentage of probability of income distribution decreases.

In order to analyse the data from each country considered, we will try to assess goodness of fit for Fermi-Dirac distribution by taking into account the minimum and the maximum value of the coefficient of determination for an annual set of data. The year with the highest coefficient of determination will be represented graphically. Of the six countries analysed, we selected the most illustrative for this kind of distribution: Finland and France. Finland has the highest coefficient of determination for the data analysed (partly due to the conversion of all values in euro 2009) and has the longest time interval analysed (1987-2009) both for mean income and for upper limit on income of disposable/net income. For the rest of the data analysed, Tables 1 and 2 present the outcomes. The graphics presented below in the following section are in log-log scale (natural logarithm).

## I.5.1. Finland

Data were provided by National Institute of Statistics from Finland [12].

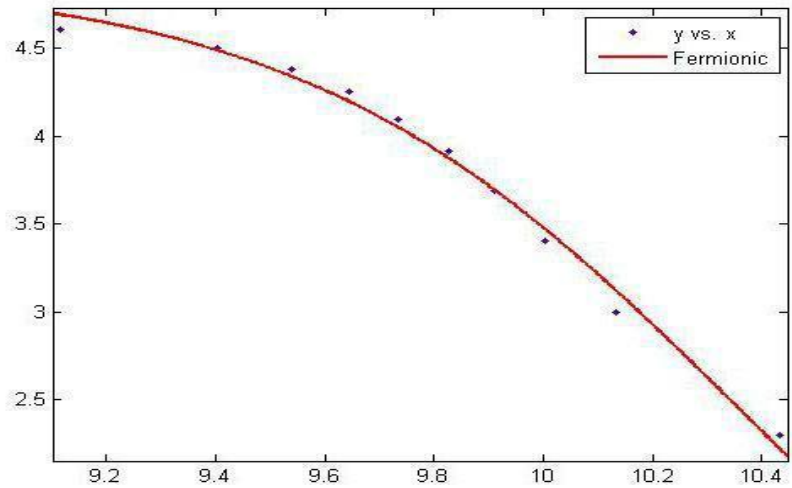

**Figure 1.** Cumulative probability distribution for mean income in Finland in the year 1990.

Characteristic parameters for the graphic were $T = 0.4007$, $c = 4.9$, $\mu = 10.36$, and $R^2 = 0.9909$. On the *x*-axis we represented logarithmic values for income deciles, while on the *y*-axis we represented logarithmic values for the cumulative probability.

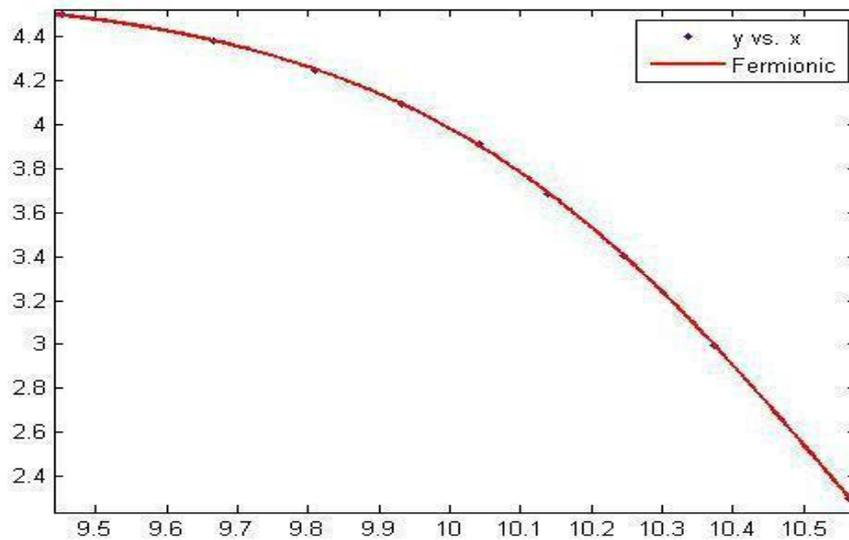

**Figure 2.** Cumulative probability distribution for upper limit on income in Finland in the year 2008.

The most illustrative analysis was considered to be the one from 1990, as the coefficient of determination was the highest for the entire period analysed. The analysis with the lowest coefficient of determination was in the year 2007, with $R^2 = 0.9755$.

Characteristic parameters for the graphic were $T = 0.3074$, $c = 4.621$, $\mu = 10.56$, $R^2 = 1$. On the *x*-axis we represented logarithmic values for income deciles while on the *y*-axis we represented logarithmic values for the cumulative probability.

The most illustrative analysis was considered to be the one from 2008, as the coefficient of determination was the highest for the entire period analysed. The analysis with the lowest coefficient of determination was in the year 1994 with $R^2 = 0.9997$.

Of all countries analysed both for mean income and upper limit on income, Finland exhibits the best fitting to the data using the fermionic distribution. A possible explanation for this is that all income figures were adjusted to the euro value from last year analysed (2009), which ensures a better representation of income unaltered by inflation or currency depreciation. Also, Finland is a country with low share of black market even compared to other developed countries. In addition, values for the coefficient of determination are higher throughout the entire period analysed for upper limit income than in the case of mean income.

### *I.5.2. France*

Data were provided by National Institute of Statistics from France [13].

Characteristic parameters for the graphic were $T = 0.3959$, $c = 4.577$, $\mu = 10.47$, and $R^2 = 1$. On the *x*-axis we represented logarithmic values for income deciles, while on the *y*-axis we represented logarithmic values for the cumulative probability.

We chose the cumulative distribution probability from the year that presented the best fit to the data regarding annual mean income before redistribution in France during 2003-2009. The lowest coefficient of determination for all annual mean income distribution was 0.9996, in the year 2008.

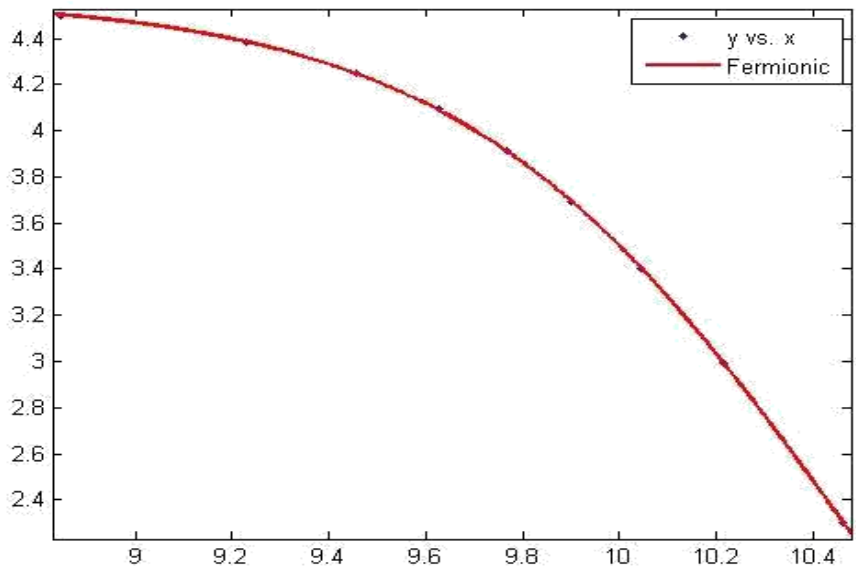

**Figure 3.** Cumulative probability distribution of mean income before redistribution for France in year 2003.

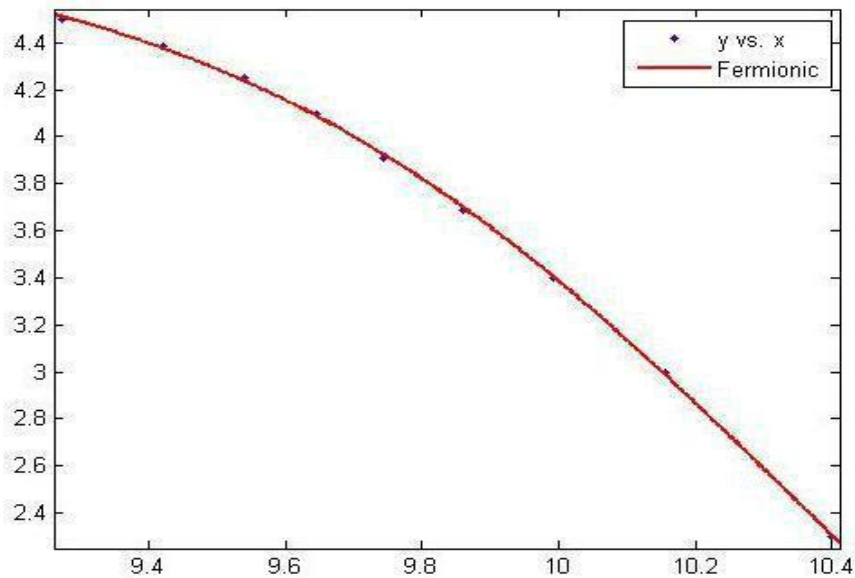

**Figure 4.** Cumulative probability distribution of inactive peoples' mean income for France in the year 2003.

Characteristic parameters for the graphic were $T = 0.4315$, $c = 4.88$, $\mu = 10.35$, and $R^2 = 0.9998$. On the $x$-axis we represented logarithmic values for income deciles while on the $y$-axis we represented logarithmic values for the cumulative probability.

We chose the cumulative distribution probability from the year that presented the best fit to the data regarding annual mean income of inactive people in France for the time interval 2002-2009. The lowest coefficient of determination for all annual mean incomes of inactive people was 0.9994 in the year 2008.

We consider this to be the most interesting finding as the income distribution among inactive individuals is not achieved thorough market mechanism but through public regulations and (in the case of France) the pensions and social benefits are managed by state administration. These two characteristics are completely different from income of the active people which are mostly are market driven. A possible explanation for this is that social aid and pensions are awarded based on the work performed by each individual and a fair contribution to the pension fund from those.

Annual values for the coefficient of determination ($R^2$) for Fermi-Dirac distribution on income distribution on population/households deciles (cumulative method) (Table 1).

**Table 1**

| Year | Upper limit on Income | | | Mean income | | | | |
|---|---|---|---|---|---|---|---|---|
| | Finland | France | Italy | Finland | France | Italy | Romania | Mexico |
| 1987 | 0.9998 | - | - | 0.9924 | - | - | - | - |
| 1988 | 0.9999 | - | - | 0.9905 | - | - | - | - |
| 1989 | 0.9998 | - | - | 0.9899 | - | - | - | - |
| 1990 | 0.9999 | - | - | 0.9909 | - | - | - | - |
| 1991 | 1 | - | - | 0.9892 | - | - | - | - |
| 1992 | 1 | - | - | 0.9874 | - | - | - | - |
| 1993 | 0.9998 | - | - | 0.9855 | - | - | - | - |
| 1994 | 0.9997 | - | - | 0.9868 | - | - | - | - |
| 1995 | 0.9999 | - | - | 0.9854 | - | - | - | - |
| 1996 | 1 | - | - | 0.9893 | - | - | - | - |
| 1997 | 0.9999 | - | - | 0.9845 | - | - | - | - |
| 1998 | 0.9999 | - | - | 0.9818 | - | - | - | - |
| 1999 | 0.9999 | - | - | 0.9746 | - | - | - | - |
| 2000 | 1 | - | 0.9986 | 0.9752 | - | 0.9927 | - | 0.9885 |
| 2001 | 0.9999 | - | - | 0.9789 | - | - | - | - |
| 2002 | 1 | 0.9998 | 0.999 | 0.9784 | - | 0.9939 | - | 0.9904 |
| 2003 | 0.9999 | 0.9998 | - | 0.9796 | 0.9853 | - | - | - |
| 2004 | 0.9999 | 0.9998 | 0.9992 | 0.9782 | 0.9833 | 0.9927 | - | 0.9874 |
| 2005 | 0.9999 | 0.9997 | - | 0.9769 | 0.9813 | - | 0.9932 | 0.9864 |
| 2006 | 1 | 0.9996 | 0.9989 | 0.9772 | 0.981 | 0.9917 | 0.9943 | 0.9896 |
| 2007 | 0.9999 | 0.9997 | - | 0.9758 | 0.9815 | - | 0.9942 | |
| 2008 | 1 | 0.999 | 0.9993 | 0.976 | 0.978 | 0.9937 | 0.9913 | 0.9892 |
| 2009 | 0.9999 | 0.9996 | - | 0.9798 | 0.9816 | - | 0.9896 | - |
| 2010 | - | - | - | - | - | - | 0.9894 | - |

Annual/monthly values for the coefficient of determination ($R^2$) for Fermi-Dirac distribution on income distribution on population/households deciles (cumulative method) (Table 2).

**Table 2**

| Year | France annual values for income before redistribution | France annual values for income of inactive people | Hong Kong monthly values for median income |
|------|------|------|------|
| 1991 | - | - | 0.9999 |
| 1996 | - | - | 0.9999 |
| 2001 | - | - | 0.9999 |
| 2002 | - | 0.9998 | - |
| 2003 | 1 | 0.9998 | - |
| 2004 | 0.9999 | 0.9998 | - |
| 2005 | 0.9999 | 0.9996 | - |
| 2006 | 0.9998 | 0.9997 | - |
| 2007 | 0.9999 | 0.9997 | - |
| 2008 | 0.9996 | 0.9994 | - |
| 2009 | 0.9999 | 0.9999 | - |

Data were provided by [12], [13], [14], [15], [16], and [17].

## I.6. Discussions and Conclusions

### *I.6.1. Conclusions regarding data analysis*

Fermionic distribution is a very good distribution regarding mean income, upper limit on income, and median income by deciles group for the countries analysed. Coefficient of determination ($R^2$) is extremely high when it comes to the fitting of data to this type of distribution. The absolute lowest value for this coefficient is 97% for the data regarding income in the countries considered. Also, compared to other distributions from relevant literature, goodness of fit provided by this type of distribution is at least similar if not better.

This paper studies for the first time the possibility of applying Statistical Mechanics methods to upper limit on income and median income by deciles group, unlike papers which consider only the distri-bution of population mean income. What is more remarkable is that Fermi-Dirac distribution has a higher coefficient of determination for the same year and country in the case of upper limit on income in comparison to the case of mean income.

In the analysis of the data considered in this paper, we tried to see goodness of fit in the case of Bose-Einstein distribution. The results yielded a lower coefficient of determination, the approximate value for most of data analysis being about 60-70%. In comparison to Fermi-Dirac distribution, in the overall value and in each particular value regarding goodness of fit to the data, Bose-Einstein distribution shows an inferior capacity to describe the distribution of population income in a country/state.

Fermi-Dirac distribution applied to income/money distribution proved to be robust especially in the case of Hong Kong and Finland data. In the case of Hong Kong, the data consisted of monthly median income (not the usual data expressed on annual bases), and the lowest coefficient determination was 0.999. Also, the Hong Kong dollar is linked to US dollar which implies a stable exchange rate and a low inflation. In case of Finland, of all the countries analysed, has the lowest share of black and grey market, which makes the data provided to be very reliable and realistic. Also, all the values are expressed in euro for the value in the year 2009.

The only exception for the scope of the application of Fermi-Dirac distribution is about the time and the countries characterised by highly inflationary processes. In our data set, this process occurs in Italy before the conversion to euro in the year 2000 and in Romania before the conversion to heavy leu in the year 2005. The total incapacity of this distribution to analyse and study these periods is not singular, as we tried to apply unsuccessfully Boltzmann-Gibbs and Bose-Einstein distributions. The main probable cause for this is that fixed earnings, which are represented by 90 % low-income part of the population, were more affected by inflation than the 10% upper income population income which is affected by the assets prices primarily. Subsequently, the share of each decile in total income changed. Also, we attempted to fit Fermi-Dirac distribution to data regarding the mean wealth distribution for France. Unfortunately, we could not fit the distribution to the data. Also, we got the same result for Bose-Einstein and Gibbs-Boltzmann distributions.

### *I.6.1.1. Fundamental causes of inequality*

In addition to the causes already identified in the literature pertaining to this subject, we would like to emphasize the whole background affecting wage distribution, not only the direct causes but some of the fundamental phenomena.

Social benefits are considered to be the main way to counterbalance the increasing inequality among population. However, social benefits are increasingly affected by the mainstream economic view regarding the relationship between national competitiveness and social welfare system. According to this, the higher the amount of funds allocated for social protection the higher is taxation level, hence less money disposable for companies to serve as a competitiveness-enhancer. The most illustrative case is the European Union, where social welfare system is unparalleled in the world. However, the EU competitiveness decreased to a large extent compared to USA and Japan, its economic contenders. Therefore, on the agenda of governments, especially the ones from the EU, there is a tendency for the roll-back of the welfare systems in order to increase the competitiveness. Subsequently, this roll-back involves a reduction in social benefits, which implies an inequality increase given that poor are the most affected by these.

One of main causes for increasing inequality is considered to be the higher capital gains of the high-income share of population. Level of taxation on capital is driven by a veritable race for capital among countries, as this is the main contributor to general welfare and prosperity. Therefore, countries tend to reduce taxation on capital by means of direct taxes on profit and other types of capital. In order to counterbalance the reduced earnings from taxes on capital, states increase labour taxes and indirect taxes such as consumption taxes, VAT for example. Subsequently, companies tend to redeploy their business in countries with lower taxes, especially capital taxes. Because of this, the high-income population earnings (which are influenced by capital gains and asset prices) had a reduction of their taxes, which in turn caused a considerable increase of their gains.

Foreign Direct Investments (FDI) affect a great deal economic development and subsequently earning structure among population. Investments abroad are an important manner for companies given the globalisation trend to increase its competitiveness. Theoretically, there are two ways to carry out with this: price competitiveness and quality competitiveness. First one deals with a decrease of costs, which allows a price reduction. The second one requires an increase of product quality [18]. In most cases, FDI occur mainly because a company can benefit from lower prices in host-country in comparison to the ones from home-country. This does not involve most often the activity with the highest technology, which is essential for quality competitiveness. In the home country, some people become unemployed which increases the wages disparity. In the host country, on the long term it creates a demand for more-skilled workers, leading to larger disparity in wages between lower-skilled and higher-skilled.

### *I.6.2. Methodological conclusions*

Fermi-Dirac distribution is very robust in describing the income distribution. Thus, we applied it successfully for households and individuals for different kinds of data (mean income, upper limit on income, and median income). Also, we applied it for annual values and mean monthly values throughout more years and sometimes for long time intervals (22 years in row in the case of Finland). More remarkably, the countries analysed had different types of economy and there are on different levels of development. For example, Finland belongs to Nordic type of economy whereas France belongs to the continental type, whereas Italy belongs to Mediterranean type. Mexico has a specific economy for Latin America, while Hong Kong is one of the Asian tigers. Romania belongs to the Eastern European type of economy, a former country in transition. Finland, France, Italy, Mexico, and Hong Kong are considered to be developed economies, while Romania is a developing economy.

We can notice higher capacity of fermionic distribution to fit the data set with upper limit on income than the data with mean income distribution. This is mainly attributable to several facts. First, the upper limit on income by deciles group describes only the income of low-income tier of the population, which accounts around 90% of the whole population. Subsequently, income of this part of population evolves in a similar manner based on the fact that the earnings of this population category are composed of wages, which are affected similarly by inflation. Second, mean income could be erroneous due to the high amount of data necessary to calculate the mean. Third, considering that earnings of the low-income tier of the population are mainly based on wages and therefore difficult to be subject for fiscal evasion, unlike the upper income tier of population (10%) for which fiscal evasion is easier to achieve. Therefore, when analysing the revenues of low-income tier of population it is advisable to use upper limit on income, which is more reliable. The best proof is the high frequency by which fermionic distribution has the coefficient of determination ($R^2$) equal to 1 (or 100%) when applied to upper limit income on deciles groups. In the case when high income tier of population is analysed or the entire income distribution is considered, the data set that should be considered for analyses is mean income.

The underlying reason for which Fermi-Dirac is a good distribution function for income is that an atomic model can be assimilated to an economy considered vertically. Thus, each stage in the chain of production can be assimilated to an atomic layer of electrons and the source of raw materials can be an analogue to nucleus. Another analogy is between added value at each stage of production and energy of each layer.

Thus, the first stage of production which corresponds to the first electrons layer has the lowest added value / lowest energy. An analogue for a product is an electron that jumps from one layer to another as it gets more energy. The more stages of productions a product undergoes, the more value added gets, hence more energy in an electrons layer further away from the nucleus. As industrial production is standardised, similar products undergo similar process of production. Even though technical process is mostly similar, general conditions regarding production factors determine different added values at each stage of production in each company for the same or similar products. Subsequently, people and companies whose revenues depend on the value added gotten at each production stage get different revenues according to company's productivity and individual productivity.

### *I.6.3. Further conclusions*

This study is the first to use (to the best knowledge of the authors) upper limit on income data in order to analyse income distribution. Also, we believe that Fermi-Dirac distribution can be used to analyse the entire regime of income distribution (both for lower-income and upper-income population). Fermi Dirac distribution proved to fit very well data and can be an alternative tool to analyse inequality.

# Chapter II. Study of correlation between thermodynamic variables and macroeconomic aggregates

## II.1. Introduction

First serious approach of the relation between Economics on one hand and Thermodynamics and Statistical Mechanics on the other hand was made by Nicholas Georgescu Roegen, which in his magnum opus [3] pointed out the possibility of using Physics as a new paradigm in the approach of Economics.

## II.2. Short Literature Review and Theoretical Background

In [11] it is dealt extensively with analogies between Thermo-dynamics and Macroeconomics. Thus, a macroeconomic system is assimilated with an isolated physical system such as a gas in a thermally isolated vessel. Assuming a state of a market with a certain number of trading agents ($N$), their assets would change over a time. It is impossible to make a detailed analysis of a market even if the initial conditions and all the forces that influence the market are known. Supposed that a huge system of interconnected equations can describe all the market forces and the solutions are found, it is very difficult (if not impossible) to use the massive amount of data in a successful manner. Instead, the alternative would be to create a statistical method based on average parameters characterising the market. This kind of average implies a vast number of microstates. Given that a market has a very large number of degrees of freedom on one hand and that many hidden and complex connections and variables may exist in the real world on the other hand, the market evolution is chaotic and unpredictable. Due to its chaotic nature, we can use in this case the ergodic hypothesis. The type of statistical ensemble used in our case is grand-canonical ensemble. Thus, total money ($M$) and number of trading agents ($N$) are not fixed, though their average is fixed for a longer for large time interval, which in our case is a year.

## II.3. Data description

The data in case of France were about individual income expressed in euro for the entire period and in yearly values both for mean income and upper limit on income by decils of population. In Italy, the data were in annual aggregated figures that comprise households' income both in lire and euro (after 2000) and both for mean income and upper limit on income by decils of households. For Finland, the data were expressed in annual aggregated figures for individuals in euro, converted to the value of euro in the final year considered (2009), both for mean income and upper limit on income by deciles of population. For Romania, the data were about annual mean income of deciles of households expressed in leu until year 2004 and afterwards in heavy leu. For Mexico, the data were about annual mean income of deciles of population (individuals), expressed in US dollars for the value in the year 2008. As for Hong Kong, the data were about mean monthly values of median income for deciles of households from three different years. The currency used was Honk Kong dollar calculated according to purchase power parity from 2001.

## II.4. Methodology

After analysing the fitted data using Fermi-Dirac distribution, we obtained thermodynamic parameters specific for such a distribution. We preferred to use for the data treatment the cumulative method. In order to study the correlation between thermodynamic variables and macro-economic aggregates, we will use only temperature and chemical potential. In tables 3 and 4, we present the outcome obtained, including the degeneracy. In the following, we present the analogies that can be drawn after the analysis of the data.

Temperature is determined by the average money per person $<m> = M/N$, where $M$ is total amount of money in a national economy and $N$ is total number of holders. The equivalent variable in Economics is called nominal income, which is represented by the amount of money a person or a household earns in a time interval, unadjusted by inflation and stated in the earned currency [19]. In the analysis of the correlation between temperature and nominal income, the normal trend for temperature is to have an overall growth considering that monetary mass and inflation increase in the long run and subsequently nominal income.

Chemical potential defined as $\mu = \dfrac{\partial E}{\partial N}\bigg|_{S,V}$ can be assimilated with the economic variable called productivity. Thus, productivity is defined as output/input ratio or (in a different manner) as the resources/inputs necessary to yield a certain output such that for the same input to get a higher output or to get the same output with a lower input. Since chemical potential is the necessary supplementary energy to add an additional number of particles into a system, it makes economic sense to assimilate it to the inverse of marginal increase of productivity. Subsequently, number of products obtained can be assimilated with newly introduced particles into the system ($\delta N$) and money required for production can be assimilated with energy required for introduction of new particles ($\delta E$) [20]. Another additional reason to assimilate chemical potential to marginal increase of productivity is caused by high correlation of productivity with income and its distribution.

However, in our case not total factor productivity matches the data regarding chemical potential but labour productivity. This is defined as:

*labour productivity =*
*= measure volume of output / measure of human input use.*

Human use can be measured as the amount of work per worker during a certain measure of time, or as results obtained per monetary unit allocated as wage and so forth [21]. Developed countries have high productivity and actually it is the most important factor that determines income level. Also, within the same country companies and people with high productivity have higher income. A good example is the income level difference between highly skilled, skilled, low-skilled, and unskilled personnel.

In the analysis of the correlation between chemical potential and labour productivity, the graphics obtained for the same country and time interval should be symmetric in reference to the *x*-axis, considering the definition above.

## II.5. Data analysis

The next step is to get annual values for temperature and chemical potential for the entire period analysed, represent them graphically, and comment the macroeconomic implications. We chose the clearest examples for correlation between temperature and nominal income on one hand and chemical potential and labour productivity on the other hand (Finland and Mexico) and the case which allows the most insightful economic analysis (France).

### II.5.1. Finland

As we can observe in figure 5, temperature/nominal income for mean income analysis has an overall increase in the time interval 1987-2007, while for 2008 and 2009 there is a sharp decrease due to the economic crisis.

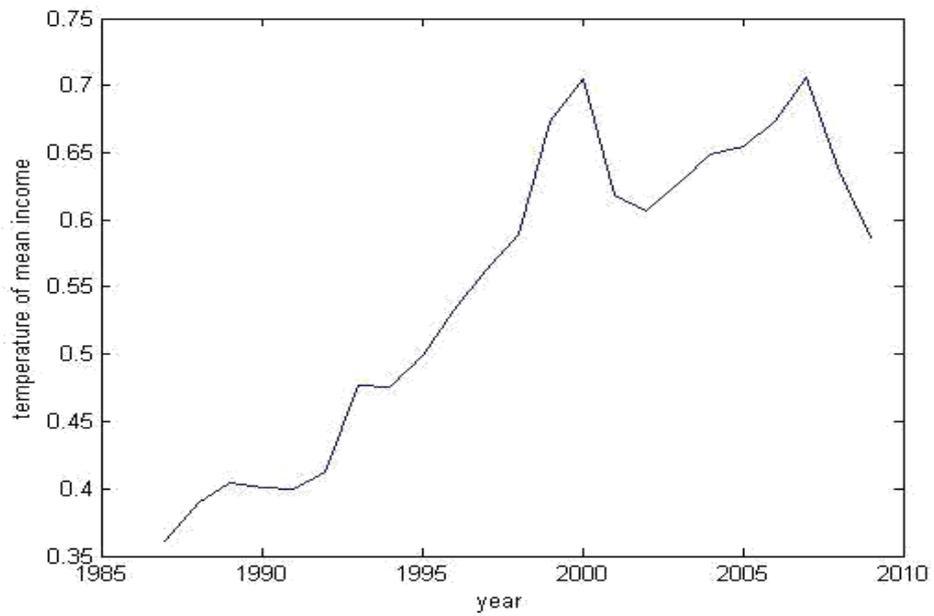

**Figure 5.** Temperature for mean income in Finland for time interval 1987-2009.

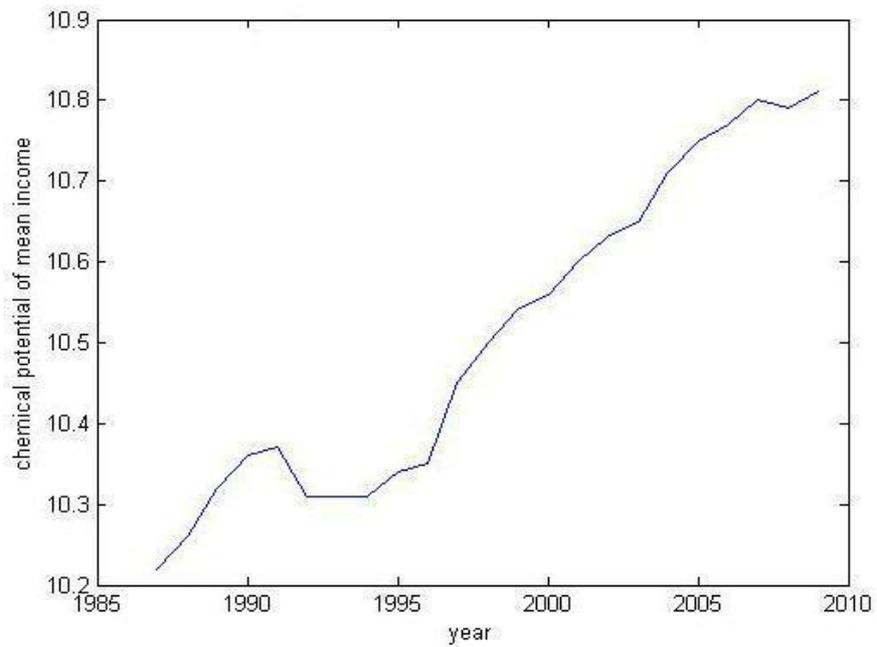

**Figure 6.** Chemical Potential for mean income in Finland for time interval 1987- 2009.

To illustrate the evolution of chemical potential/labour productivity as presented in figure 6, in the opinion of Finish National Institute of Statistics "pace of growth in labour productivity has slowed down strongly in the whole economy since the mid-1990s, from 3.5 per cent to 0.7 per cent in 2010. Although the annual growth rates for individual years (2004, 2007 and 2010) have been around 3 per cent the trend in labour productivity has been declining since the mid-1990s" [22].

Figure 7 illustrates the increase of nominal income that occurred as a consequence of the economic recovery beginning with the year 1995. From the year 2007 on there is a sharp decrease in the nominal income due to the crisis. We can notice that its overall tendency follows the same distribution as in the case of mean income. The exception is that for the time interval 2000-2005, when the increase of temperature is not as big and accentuated as in case on the analysis performed on mean income set of data.

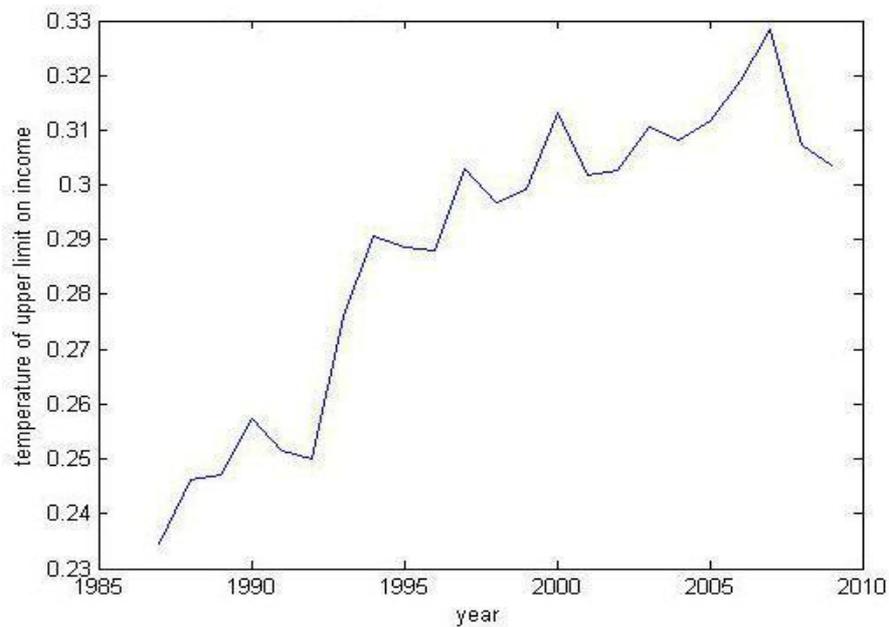

**Figure 7**. Temperature for upper limit on income in Finland for time interval 1987-2009.

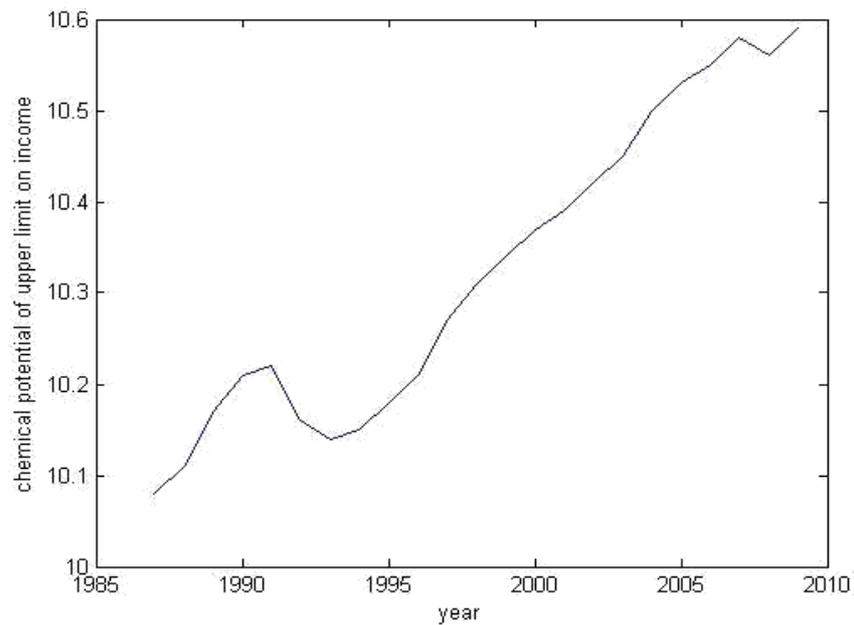

**Figure 8.** Chemical Potential for upper limit income in Finland
for time interval 1987-2009.

From figure 8 we can notice that chemical potential in the case of upper limit income exhibits the same trends as in the case of mean income.

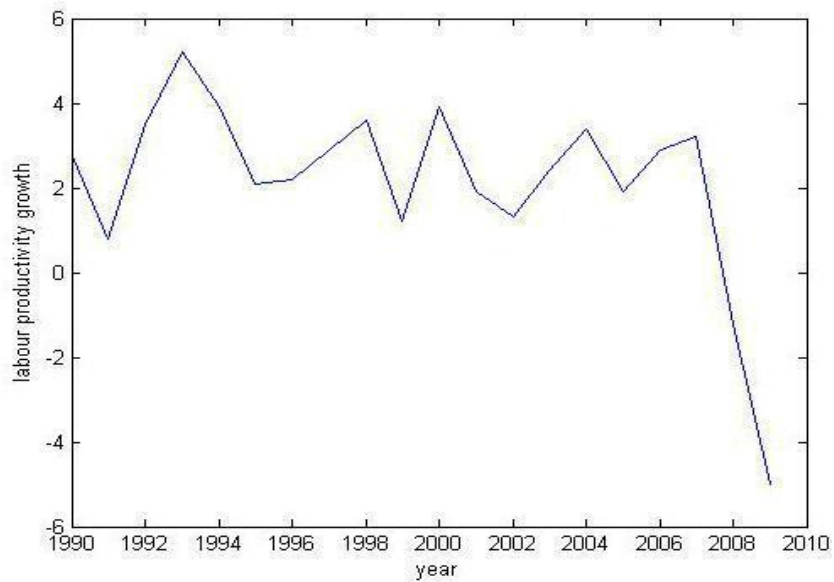

**Figure 9.** Labour productivity growth rate in Finland
for time interval 1990-2009.

Data were provided by [23]. In figure 9 we displayed the graphical tendencies of labour productivity annual growth rate as proxy for chemical potential. Considering that chemical potential according to its definition, it is fairly symmetrical to its proxy about x axis. We can observe that main trends of labour productivity annual growth rate are captured in the evolution of the chemical potential. Subsequently, we can observe the better evolution of productivity in the middle of the 90s and slower growth in the aftermath, and the sharp decrease at the same time with the commencement of the crisis in the year 2007.

We can conclude after having a look at the overall trends of chemical potential and temperature that nominal income increased while productivity growth rate was slower. In order to have a better picture of the Finnish economy we would need the evolution of inflation, but these trends are not a good indicator for economic activity, especially for its national competitiveness.

### II.5.2. France

In figure 10, temperature (which indicates the evolution of nominal income i.e. the physical amount of money that a person obtains) shows a normal behaviour except the slight decrease in the year 2004. In the 2008, when in times of deep economic crisis the nominal income increases. However, we must take into account that measures taken in order counterbalance the effects of the crisis did not affect the income in France to the same extent as was the case in other developed countries. However, we can notice that for the time interval 2003-2008 (before the economic crisis), temperature/nominal income presents an overall increase.

In figure 11, the evolution of chemical potential is in full agreement with the previous results for France. Thus, we notice the its decrease up to the "peak" of productivity during 2004, followed by an increase unaffected significantly by the crisis and from 2009 there are the first signs of a small recovery presented by an slighter increase than in the case of the previous two years. Chemical potential obtained from both sets of data regarding mean income and upper limit on net income on one hand and mean income before redistribution on the other hand show remarkable similarities.

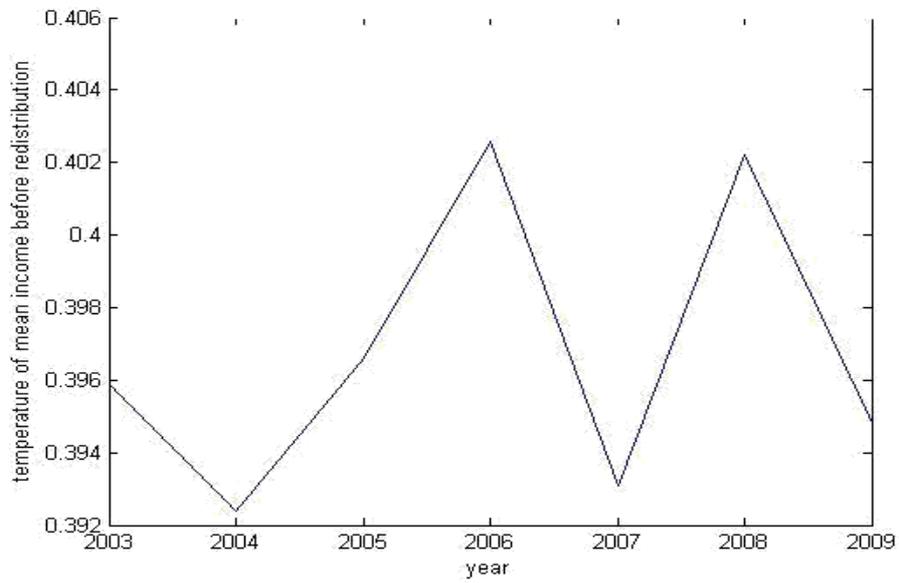

**Figure 10.** Temperature for France income before redistribution
for time interval 2003-2009.

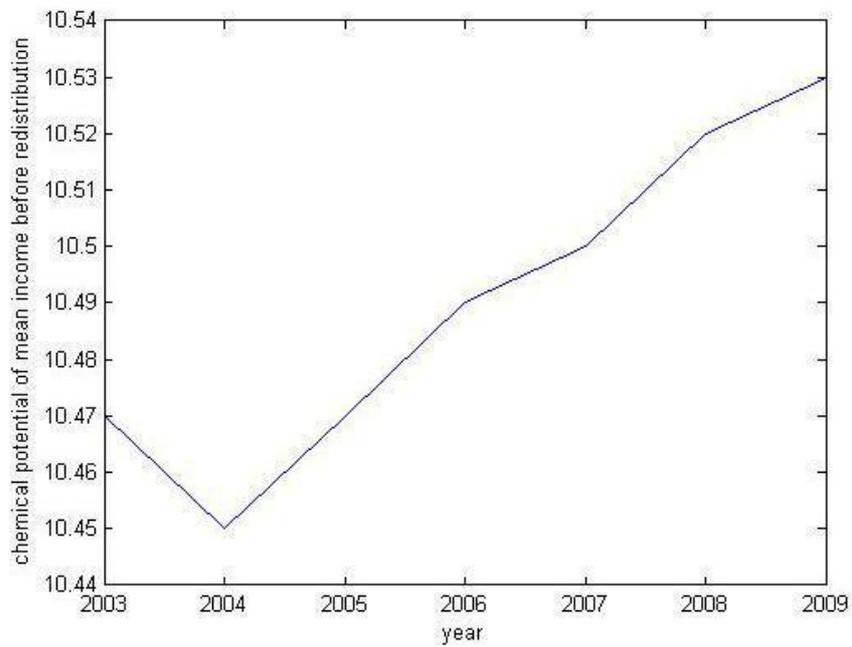

**Figure 11.** Chemical Potential for France income before distribution
for time interval 2003-2009.

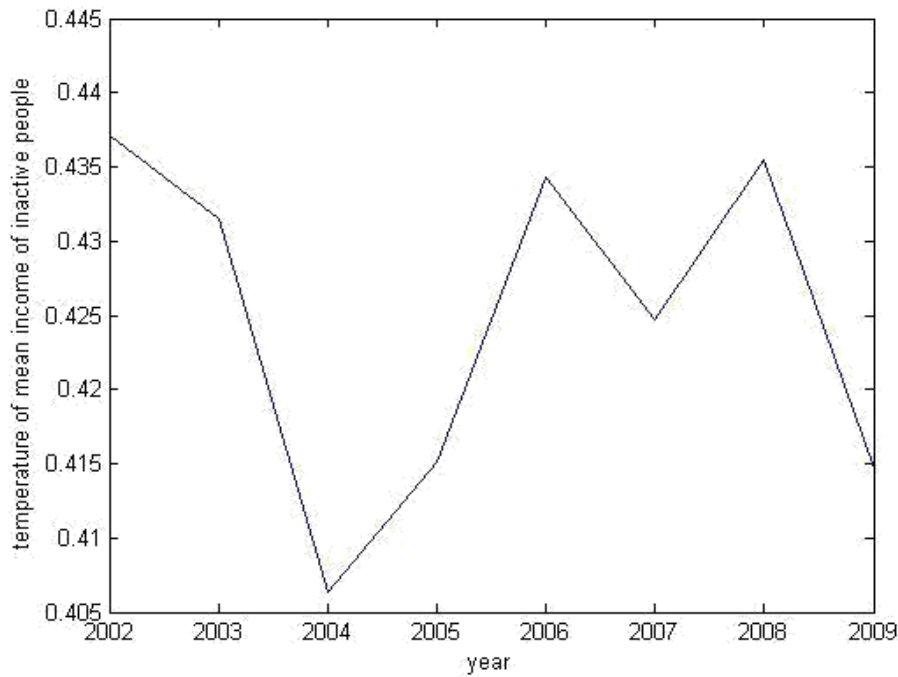

**Figure 12.** Temperature for France mean income of inactive people for time interval 2002-2009.

In figure 12, temperature follows the same trends as the temperature extracted from data about mean income before distribution. This implies that is a remarkable good similarity between the evolution income before distribution and the income for inactive people, which consists mainly of pensions and social benefits. The evolution in the year 2007 for temperature both for income before redistribution and inactive people's income highlights once again the tight correlation between gross income and various social benefits not only as general trend but for specific time intervals as well.

In figure 13, the evolution of chemical potential of mean income of inactive people is in full accordance with the ones from mean income, upper limit on income, and mean income before redistribution. The evolution of chemical potential shows the above mentioned trend even more than in the case of evolution of temperature.

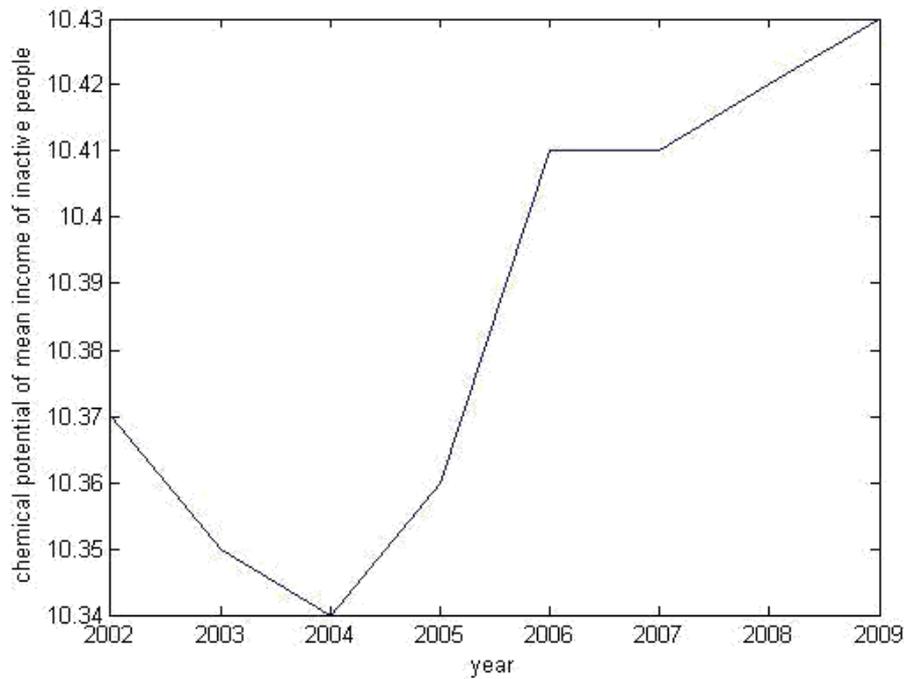

**Figure 13.** Chemical Potential for mean income of inactive people
for the time interval 2002-2009.

### II.5.3. Mexico

In figure 14, Temperature indicates a normal increase of nominal income for years 2005 and 2006 followed afterwards by a decrease caused by the crisis. However, there is an abnormality regarding the decrease of temperature (nominal income) in the years 2002 and 2004, while for the next years the general level stays under the level from the year 2000. A possible explanation is that during the time interval 2000-2002 the labour productivity had small increments or negative increments/drops, when it is known that productivity is one of economic variables that determines to a very large extent the income level.

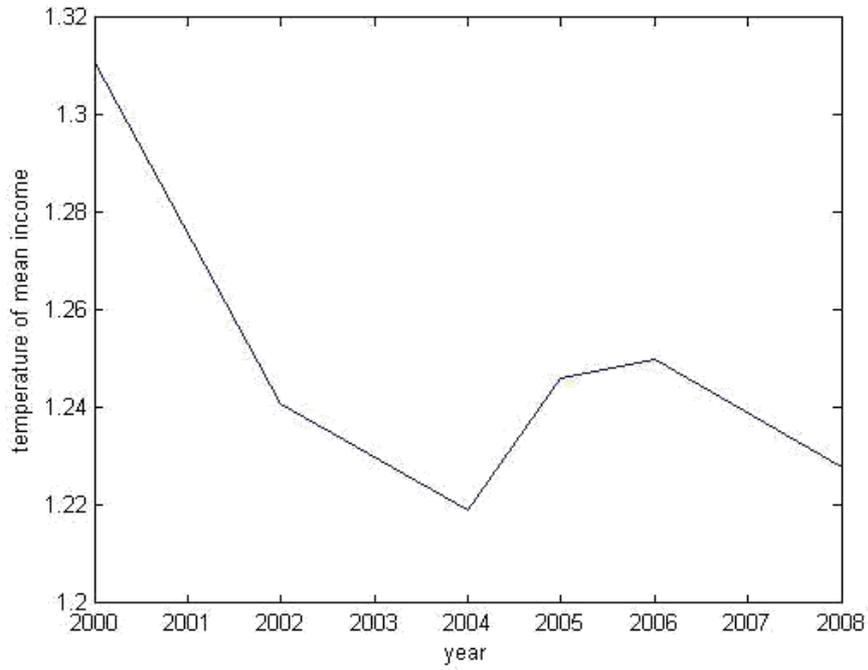

**Figure 14.** Temperature for mean income in Mexico for time interval 2000-2008.

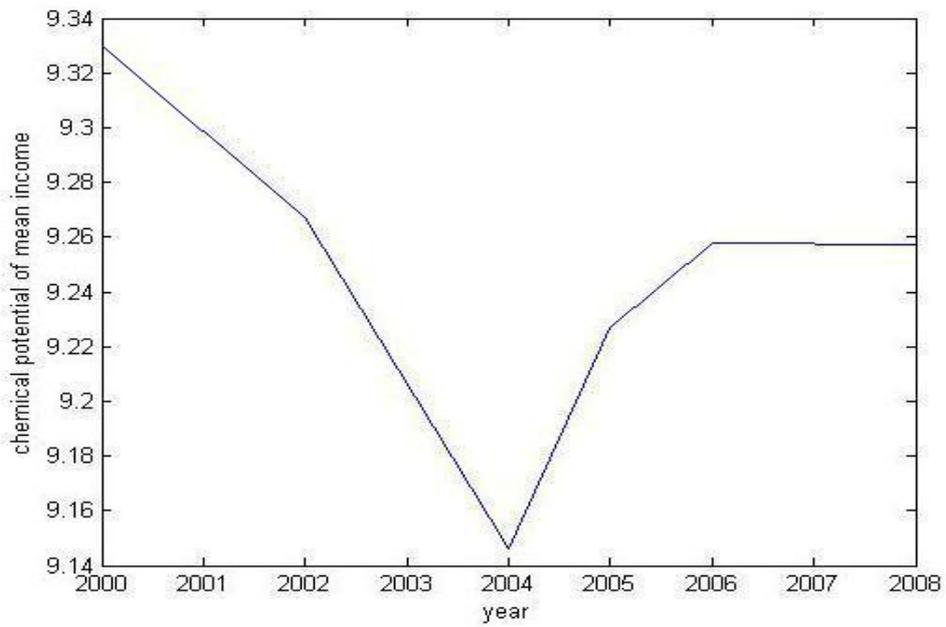

**Figure 15.** Chemical Potential for mean income in
Mexico for the time interval 2000-2008.

In figure 15, The evolution of chemical potential indicates a "peak" of labour productivity in the year 2004, followed by slower growth of productivity until the year 2006. Afterwards, during the crisis, productivity growth succeeded to stay relatively normal. A possible explanation for this is that Mexican companies succeeded to cut personnel costs.

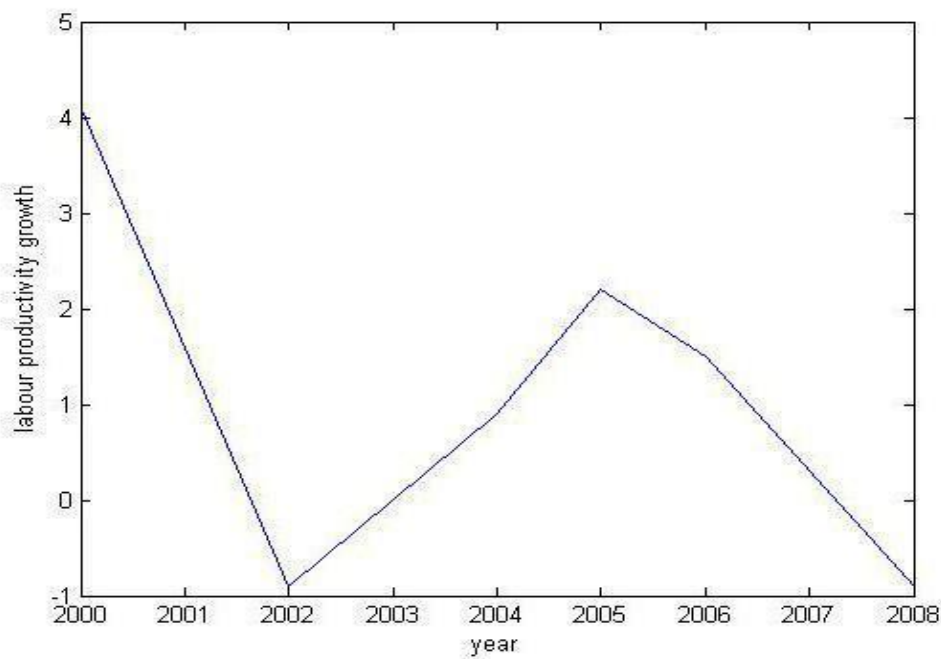

**Figure 16.** Labour productivity annual growth rate in Mexico for the time interval 2000-2008.

Data were provided by [23]. In figure 16, we displayed the graphical tendencies of labour productivity annual growth rate in Mexico as proxy for chemical potential. Considering that chemical potential according to its definition is fairly symmetrical to its proxy about x axis, one can observe that main trends of labour productivity annual growth rate are captured in the evolution of the chemical potential. Subsequently, the peak of the evolution of labour productivity growth in the year 2004 and its decrease starting from the year 2006 on are highlighted. However, the increase from the year 2005 of labour productivity is not captured in the evolution of the chemical potential.

Annual parameters of Fermionic distribution applied to income distribution by deciles of population/households (cumulative method) (Table 3).

**Table 3**

| Country | Year | Upper limit on income | | | Mean income | | |
|---|---|---|---|---|---|---|---|
| | | $T$ | $C$ | $\mu$ | $T$ | $C$ | $\mu$ |
| Finland | | | | | | | |
| | 1987 | 0.2344 | 4.588 | 10.08 | 0.361 | 4.827 | 10.22 |
| | 1988 | 0.2461 | 4.609 | 10.11 | 0.3893 | 4.874 | 10.26 |
| | 1989 | 0.2472 | 4.602 | 10.17 | 0.4048 | 4.899 | 10.32 |
| | 1990 | 0.2573 | 4.633 | 10.21 | 0.4007 | 4.9 | 10.36 |
| | 1991 | 0.2516 | 4.632 | 10.22 | 0.3993 | 4.902 | 10.37 |
| | 1992 | 0.2501 | 4.645 | 10.16 | 0.413 | 4.955 | 10.31 |
| | 1993 | 0.276 | 4.692 | 10.14 | 0.477 | 5.065 | 10.31 |
| | 1994 | 0.2906 | 4.725 | 10.15 | 0.4762 | 5.064 | 10.31 |
| | 1995 | 0.2886 | 4.702 | 10.18 | 0.4989 | 5.09 | 10.34 |
| | 1996 | 0.2881 | 4.671 | 10.21 | 0.5332 | 5.205 | 10.35 |
| | 1997 | 0.3029 | 4.674 | 10.27 | 0.5624 | 5.14 | 10.45 |
| | 1998 | 0.2967 | 4.636 | 10.31 | 0.5893 | 5.147 | 10.5 |
| | 1999 | 0.2992 | 4.64 | 10.34 | 0.6739 | 5.314 | 10.54 |
| | 2000 | 0.3133 | 4.65 | 10.37 | 0.7052 | 5.349 | 10.56 |
| | 2001 | 0.3018 | 4.63 | 10.39 | 0.6181 | 5.159 | 10.6 |
| | 2002 | 0.3028 | 4.633 | 10.42 | 0.6066 | 5.135 | 10.63 |
| | 2003 | 0.3106 | 4.643 | 10.45 | 0.6282 | 5.175 | 10.65 |
| | 2004 | 0.3083 | 4.625 | 10.5 | 0.6489 | 5.187 | 10.71 |
| | 2005 | 0.3117 | 4.631 | 10.53 | 0.6543 | 5.196 | 10.75 |
| | 2006 | 0.3191 | 4.633 | 10.55 | 0.6734 | 5.212 | 10.77 |
| | 2007 | 0.3283 | 4.642 | 10.58 | 0.7062 | 5.247 | 10.8 |
| | 2008 | 0.3074 | 4.621 | 10.56 | 0.6354 | 5.135 | 10.79 |
| | 2009 | 0.3036 | 4.618 | 10.59 | 0.5873 | 5.066 | 10.81 |
| France | 2002 | 0.3946 | 4.734 | 10.4 | - | - | - |
| | 2003 | 0.3835 | 4.721 | 10.39 | 0.6644 | 5.134 | 10.59 |
| | 2004 | 0.3745 | 4.711 | 10.38 | 0.6793 | 5.17 | 10.58 |
| | 2005 | 0.3778 | 4.712 | 10.39 | 0.673 | 5.121 | 10.62 |
| | 2006 | 0.3906 | 4.73 | 10.42 | 0.7019 | 5.165 | 10.64 |
| | 2007 | 0.3824 | 4.713 | 10.44 | 0.6904 | 5.144 | 10.66 |
| | 2008 | 0.3901 | 4.746 | 10.45 | 0.7074 | 5.185 | 10.67 |
| | 2009 | 0.387 | 4.716 | 10.46 | 0.6796 | 5.112 | 10.69 |
| Italy | 2000 | 0.4358 | 4.566 | 10.77 | 0.6885 | 4.821 | 11.04 |
| | 2002 | 0.4421 | 4.573 | 10.84 | 0.6966 | 4.839 | 11.1 |
| | 2004 | 0.4607 | 4.623 | 10.88 | 0.7323 | 4.924 | 11.13 |
| | 2006 | 0.4254 | 4.59 | 10.93 | 0.7266 | 4.938 | 11.19 |
| | 2008 | 0.4616 | 4.616 | 10.98 | 0.7111 | 4.886 | 11.23 |
| Romania | 2005 | - | - | - | 0.7977 | 5.722 | 7.581 |
| | 2006 | - | - | - | 0.7926 | 5.547 | 7.787 |
| | 2007 | - | - | - | 0.7419 | 5.461 | 7.991 |
| | 2008 | - | - | - | 0.6739 | 5.355 | 8.228 |
| | 2009 | - | - | - | 0.6382 | 5.385 | 8.274 |
| | 2010 | - | - | - | 0.6245 | 5.468 | 8.227 |
| Mexico | 2000 | - | - | - | 1.311 | 5.102 | 9.33 |
| | 2002 | - | - | - | 1.241 | 5.088 | 9.267 |
| | 2004 | - | - | - | 1.219 | 5.076 | 9.146 |
| | 2005 | - | - | - | 1.246 | 5.094 | 9.227 |
| | 2006 | - | - | - | 1.25 | 5.141 | 9.258 |
| | 2008 | - | - | - | 1.228 | 5.086 | 9.257 |

Annual/monthly parameters of Fermionic distribution applied to income distribution by deciles of population/households (cumulative method) (Table 4).

**Table 4**

| Year | France annual income before redistribution | | | France annual income of inactive peoples | | | Hong Kong monthly median income | | |
|---|---|---|---|---|---|---|---|---|---|
| | T | C | μ | T | C | μ | T | C | μ |
| 1991 | - | - | - | - | - | - | 0.6161 | 4.654 | 10.22 |
| 1996 | - | - | - | - | - | - | 0.615 | 4.638 | 10.79 |
| 2001 | - | - | - | - | - | - | 0.6188 | 4.587 | 10.92 |
| 2002 | - | - | - | 0.4372 | 4.871 | 10.37 | - | - | - |
| 2003 | 0.3959 | 4.577 | 10.47 | 0.4315 | 4.88 | 10.35 | - | - | - |
| 2004 | 0.3924 | 4.582 | 10.45 | 0.4064 | 4.839 | 10.34 | - | - | - |
| 2005 | 0.3966 | 4.585 | 10.47 | 0.4151 | 4.842 | 10.36 | - | - | - |
| 2006 | 0.4026 | 4.592 | 10.49 | 0.4343 | 4.854 | 10.41 | - | - | - |
| 2007 | 0.3931 | 4.578 | 10.5 | 0.4247 | 4.846 | 10.41 | - | - | - |
| 2008 | 0.4022 | 4.599 | 10.52 | 0.4355 | 4.88 | 10.42 | - | - | - |
| 2009 | 0.3948 | 4.578 | 10.53 | 0.4146 | 4.824 | 10.43 | - | - | - |

## II.6. Conclusions

### II.6.1. Conclusions regarding the data analysis

Temperature exhibited a relative overall trend of increase as the amount of money normally increases. For most of these countries, the year 2007 marked the beginning of crisis and therefore temperature/nominal income decreased. Exception was for Mexico in year 2006 and Romania in 2008.

Chemical potential was supported by data regarding labour produc-tivity in its overall trends. For Mexico, the chemical potential represen-tation captures the peak of labour productivity growth in 2004 and afterwards the downfall corresponding to the crisis from 2006. Finland was characterised by a bigger growth of labour productivity in the 90s followed by a dramatic decrease in years 2000.

France showed a remarkable similarity regarding temperature and chemical potential between across all three sets of data (gross income, net income, and inactive people income) which indicates also a very good reliability of data.

### II.6.2. Methodological conclusions

First methodological consequence is that study of income distribution is useful to analyse social system and taxation system. In the case of France was very useful to compare the performed analysis on income before redistribution, net income, and social benefits. We can notice that results in the case of income before redistribution and net income are very similar. This indicates that taxation has a high degree of fairness, as no decile of population's income is overtaxed. Also, if we examine the shape of income distribution between net income and inactive people, we observe that they are very similar. As a consequence, we can say that this shows that social welfare system is very well linked to the general trends of income and productivity. We can state also that social benefits and pensions are awarded considering the economic efficiency and no electoral gifts were noticed during electoral years. Thus, there is a genuine concern that social benefits to be awarded such that they do not harm the macroeconomic equilibrium.

In the analysis of temperature, some graphics exhibit (the most illustrative case is Mexico) an abnormal evolution of temperature for few years, when there is a drop. Thus, nominal income should not decrease except times of deep recession, as was the case from 2007 for most countries analysed. This situation could be explained by several phenomena such as an increased level of taxation (in case of net income), capital withdrawal, restrictions regarding credits imposed by National Bank, reduction of money quantity by National Bank. Another explanation for the decrease of temperature/nominal income is represented by an increase of number of unemployed. Thus, this part of population has no wage or (only) social benefits (which are traditionally low) and as a result the amount of money per capita drops. However, there is necessary a more in depth analysis, especially an econometric study. A possible solution is to consider temperature as an analogue to an index which would contain other economic variables such as GDP.

Chemical potential was found to be a good overall proxy for labour productivity. Analyses for on countries which we not included in this study support this conclusion as well. For instance, the evolution of labour productivity in the USA in the year 2002 is in conformity with the evolution of the chemical potential [11].

More in depth analysis of the chemical potential should be performed, especially by using econometric methods.

### *II.6.3. Further conclusions*

Further analysis could be done in order to fully explain the underlying phenomena regarding chemical potential. Next step would be to consider the activity coefficient which is related to both chemical potential and temperature and we believe it to describe overall profitability.